\documentclass[conference]{IEEEtran}
\IEEEoverridecommandlockouts

\usepackage{color}
\usepackage{multirow}
\usepackage{booktabs}
\usepackage{amsmath,graphicx,url,times,booktabs,tabularx,amsfonts,siunitx}
\usepackage{algorithmic}
\usepackage{setspace}
\usepackage{graphicx}
\usepackage{textcomp}
\usepackage{xcolor}
\usepackage{latexsym}
\usepackage{bbding}
\usepackage{footnote}
\usepackage{arydshln} % 导入绘制虚线的宏包
\usepackage{hyperref}
\usepackage{cite}
\usepackage{amsmath,amssymb,amsfonts}
\usepackage{algorithmic}
\usepackage{graphicx}
\usepackage{textcomp}
\usepackage{xcolor}

\def\BibTeX{{\rm B\kern-.05em{\sc i\kern-.025em b}\kern-.08em
    T\kern-.1667em\lower.7ex\hbox{E}\kern-.125emX}}

\title{Sound-VECaps: Improving Audio Generation with Visually Enhanced Captions}

\author{
    \IEEEauthorblockN{
    Yi Yuan\IEEEauthorrefmark{2}, 
    Dongya Jia\IEEEauthorrefmark{3}, 
    Xiaobin Zhuang\IEEEauthorrefmark{3}, 
    Yuanzhe Chen\IEEEauthorrefmark{3},
    Zhuo Chen\IEEEauthorrefmark{3},
    Yuping Wang\IEEEauthorrefmark{3},
    Yuxuan Wang\IEEEauthorrefmark{3},\\
    Xubo Liu\IEEEauthorrefmark{2}, 
    Xiyuan Kang\IEEEauthorrefmark{2},
    Mark D. Plumbley\IEEEauthorrefmark{2}, 
    Wenwu Wang\IEEEauthorrefmark{2}
}
\IEEEauthorblockA{
    \IEEEauthorrefmark{2}Centre for Vision, Speech and Signal Processing (CVSSP), University of Surrey
    \IEEEauthorrefmark{3}ByteDance \\
    }
}

\begin{document}

\maketitle

\begin{abstract}
Generative models have shown significant achievements in audio generation tasks. However, existing models struggle with complex and detailed prompts, leading to potential performance degradation. We hypothesize that this problem stems from the simplicity and scarcity of the training data. This work aims to create a large-scale audio dataset with rich captions for improving audio generation models. We first develop an automated pipeline to generate detailed captions by transforming predicted visual captions, audio captions, and tagging labels into comprehensive descriptions using a Large Language Model~(LLM). The resulting dataset, Sound-VECaps, comprises $1.66$M high-quality audio-caption pairs with enriched details including audio event orders, occurred places and environment information. We then demonstrate that training the text-to-audio generation models with Sound-VECaps significantly improves the performance on complex prompts. Furthermore, we conduct ablation studies of the models on several downstream audio-language tasks, showing the potential of Sound-VECaps in advancing audio-text representation learning. Details of our dataset and demos are available \href{https://yyua8222.github.io/Sound-VECaps-demo/}{here}. 

\end{abstract}

\begin{IEEEkeywords}
Audio generation, audio retrieval, diffusion model, audio-language dataset
\end{IEEEkeywords}

\section{Introduction}
\label{sec:intro}
Generative models have recently achieved substantial success for text-to-audio generation. In particular, the development of language models~\cite{clap,t5} and diffusion models~\cite{dalle3,stable_diffusion} have enabled the creation of powerful systems~\cite{audioldm2,tango} on generating high-fidelity audio clips. 

Despite their success in generating audio with simple captions, current models struggle with complex prompts containing detailed information, which referred to the challenge as ``prompt following''~\cite{dalle3}. A potential reason for this limitation is that existing audio-caption datasets often lack in quantity and quality~(detailed information) of the captions. In most of these datasets, each audio is matched with simple and short captions, typically, fewer than $10$ words. As a result, the captions in these datasets may not contain fine-grained information that could be useful for highly controllable audio generation. 

In addition, the simplicity of the caption often results in situations where the same caption corresponds to multiple audio files~(e.g., there are $2.5$K audio clips match with the caption `` Music is playing'' in WavCaps~\cite{wavcaps}), causing the system to avoid learning specific audio feature and lead to more instability in the generated outputs. A possible way to address this issue is to incorporate additional information, such as visual features, which have been shown to provide more detailed insights. One of the previous attempts is the Auto-ACD~\cite{autoacd}, where video features are used to improve the description of the event-occurring scene. However, Auto-ACD only takes the visual feature of the middle frame, and the caption has been designed to ignore the visual-only contents, losing more detailed information. 

\begin{table}[t]
\caption{The analysis of audio-caption datasets. \textbf{Loc} and \textbf{Env} are the captions including locations and environmental information.}
\centering
\small
\resizebox{0.475\textwidth}{!}{%

\begin{tabular}{ccccc}
\toprule
{\textbf{Dataset}}  & \multicolumn{1}{c}{\textbf{Number}} & \multicolumn{1}{c}{\textbf{Avg. Len}} & \multicolumn{1}{c}{\textbf{Loc. Inf}} & \multicolumn{1}{c}{\textbf{Env. Inf}} \\
\midrule
AudioSet~\cite{audioset} &$2.1$M & $3$ &Label& Label \\

Clotho~\cite{clotho} &$5$K &  $11$ & $1.2$K & $0.9$K \\

AudioCaps~\cite{audiocaps} &$46$K &  $9$ &  $4$K & $3$K \\

WavCaps~\cite{wavcaps} &$400$K &  $8$ & $51$K  &$37$K  \\

Auto-ACD~\cite{autoacd} &$1.9$M &  $18$ & $1.23$M & $69$K \\
\midrule
Sound-VECaps$_{\operatorname{A}}$ & $1.66$M & $31$ &$1.44$M & $1.36$M \\

Sound-VECaps$_{\operatorname{F}}$ & $1.66$M & $40$ &$1.46$M &
$1.38$M\\

\bottomrule
\end{tabular}
}

\label{tab:datasize}
\vspace{-0.3cm} % 减少0.5厘米的垂直空间
\end{table}

In this paper, we aim to leverage external visual guidance to enhance the audio captions. With improved captions, we can provide better alignment between the prompt and the sound, thereby improving text-to-audio generation systems. Specifically, we propose new pipelines to construct a large-scale audio-language dataset with vision-enhanced captions. Our approach first involves collecting external visual information using state-of-the-art~(SoTA) image captioning models. These visual captions, combined with simple audio information, are then used to create new, enriched captions through Large Language Models (LLMs). By incorporating additional visual information, our method ensures the accuracy of audio details while enhancing the captions with comprehensive content, including temporal, spatial, and contextual elements related to the environment. Building on AudioSet~\cite{audioset}, we introduce Sound-VECaps, a large-scale dataset comprising over $1.66$M audio-caption pairs. 

%This dataset leverages image captions to produce more detailed and contextually rich audio captions.

\begin{figure*}[htbp]
    \centering
\includegraphics[width=1\linewidth]{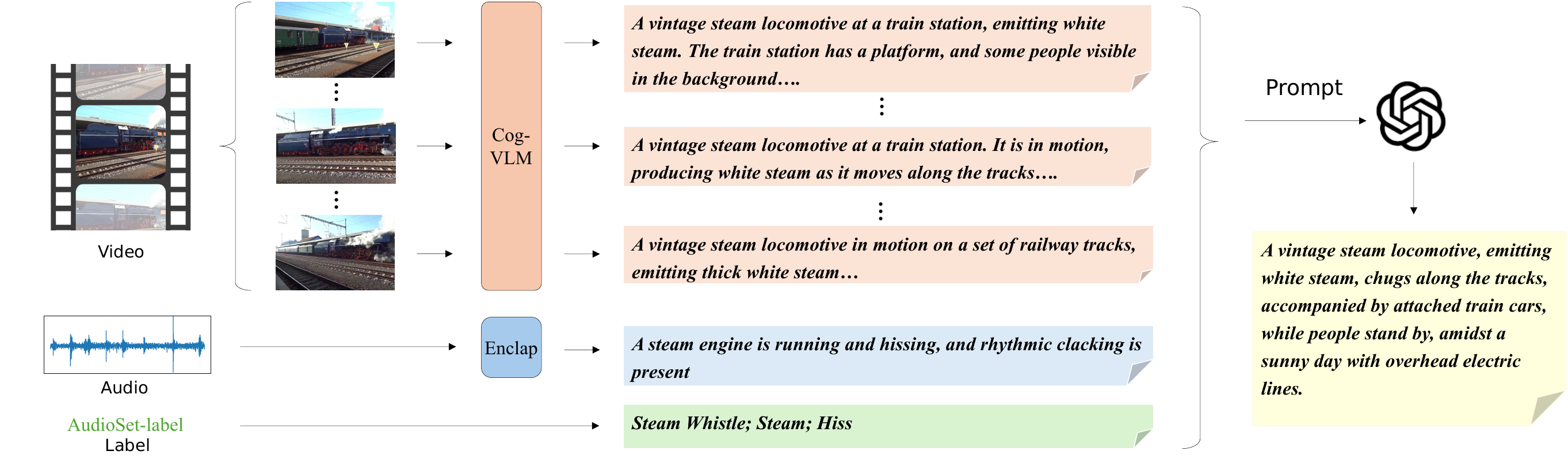}
    \caption{The caption generation pipeline of the Sound-VECaps}
    \label{fig:overview}
\vspace{-0.3cm} % 减少0.5厘米的垂直空间
\end{figure*}

Using Sound-VECaps as the training dataset, our experiments with the audio generation model, AudioLDM~\cite{audioldm}, show substantial improvements over baseline models. To evaluate the performance on complex and extended prompts, we propose a new benchmark for text-to-audio generation by constructing an enhanced AudioCaps~\cite{audiocaps} testing set~(same audio with better captions) named AudioCaps-Enhanced. Specifically, the AudioLDM-Large trained on Sound-VECaps achieves a Frechet Audio Distance~(FAD) score of $1.49$ on the AudioCaps. It further improves to a score of $1.06$ on AudioCaps-Enhanced, significantly outperforming current SoTA models. Moreover, we conduct experiments on Sound-VECaps across various audio-language tasks, demonstrating that systems trained on Sound-VECaps achieve SoTA performance in specific audio-domain tasks, such as audio retrieval. We also investigate the effectiveness of the visual-only content within the caption and the impact of these features during inference. In addition, an external version of Sound-VECaps that excludes all the visual-only information~(Sound-VECaps$_{\operatorname{A}}$) is also provided for different research purposes.   

% This paper is organized as follows. Section~\ref{sec:Dataset Construction} introduces the dataset, followed by the audio generation system trained on our datasets in Section~\ref{sec:method}. Section \ref{sec:exp} presents the experimental results and conclusions are given in Section \ref{sec:conclusion}.

\section{Audio Dataset}
\label{sec:Dataset Construction}

Our Sound-VECaps dataset is built on AudioSet~\cite{audioset} with the processing pipeline shown in Figure~\ref{fig:overview}. In particular, LLMs are prompted to generate captions based on three pieces of text information, namely, visual captions from the video, audio captions from the waveform, and the label tags in the original dataset. 

\subsection{Captions from Video}
One of the novel aspects of the proposed dataset is that we leverage the caption of the corresponding video to provide more detailed information about the audio events. Different from previous visual-related approaches~\cite {autoacd} that only apply the visual information of middle frames, the proposed strategy utilizes the captions of complete videos to secure more detailed descriptions. On the other hand, current SoTA video captioning systems~\cite{valor,plug2} mainly pool all the visual information into an aligned feature dimension, losing the temporal information~(order of the events). Hence, we capture visual information for multiple frames through image captioning to maintain this temporal information. Specifically, we follow the image caption generation of Stable Diffusion 3~\cite{stable-diffusion-3} and apply the CogVLM~\cite{cogvlm} captioning system. To improve the efficiency, the system only captions the frame of each video by second e.g., $11$ captions for a $10$-second audio.

\subsection{Captions from Audio}
We found that captions directly from video sometimes may not reflect the correct events, such as background and invisible sound. Hence, two constraints are provided to guide the LLM to understand the auditory information: the label provided by the original AudioSet dataset, and the simple audio caption generated by audio captioning models. We apply the SoTA captioning model, EnCLAP~\cite{enclap}, to generate concise and brief captions of each audio clip. 

\subsection{Proposed Caption Generation}
Combining three textual information mentioned above, an LLM is applied to generate the final caption, where we use Llama3-7B~\cite{llama} to assemble re-caption the comprehensive description of each audio. 

\subsection{Dataset Processing}
Due to the issues of some videos being too old~(not accessible anymore), we collected a total of $1.81$M videos from the AudioSet. In addition, around $10$k video clips are skipped due to the sensitive policy of the LLMs~(e.g., violence). Furthermore, we found that some video clips present static visual information with complete background sounds, leading the caption focusing on visual events but ignoring the actual audio events. To ensure the correctness of the visual guidance and improve the data quality, a filtering strategy is applied to detect and exclude the captions of static video which presents more than $80$\% same frames. Overall, we obtain the Sound-VECaps datasets containing $1.66$M audio-caption pairs. The Sound-VECaps provides two different versions of captions for various purposes, specifically, Sound-VECaps$_{\operatorname{A}}$ removes visual-only information and contains only audible contents or environmental-descriptive information, while Sound-VECaps$_{\operatorname{F}}$ describes full detailed information including visual features, e.g., texts, names, shapes, and colours.

% \subsection{Dataset Statistics}

\begin{table*}[htbp]
\caption{The comparison between generation frameworks evaluated on AudioCaps~(previous benchmarks) and AudioCaps-Enhanced~(proposed benchmarks). Both $\operatorname{CLAP}_{\operatorname{score}}$(\%) and MOS are only evaluated on the best results of each system. AC and AS are short for AudioCaps~\cite{audiocaps} and AudioSet~\cite{audioset} respectively.}
\centering
\small
\begin{tabular}{ccccc:ccc:cc}
\toprule
                     \multirow{3}{*}{\textbf{Model}}  & \multirow{3}{*}{\textbf{Training Dataset}} & \multicolumn{3}{c}{\textbf{AudioCaps}} &  \multicolumn{3}{c}{\textbf{AudioCaps-Enhanced}} & \multicolumn{2}{c}{\textbf{Best Result}}\\
 \cmidrule(lr){3-10}
 & & KL $\downarrow$ &  IS $\uparrow$& FAD $\downarrow$ & KL $\downarrow$ &  IS $\uparrow$& FAD $\downarrow$ & $\operatorname{CLAP}_{\operatorname{score}}$(\%)$\uparrow$ &  MOS$\uparrow$ \\
\midrule
AudioGen~\cite{audiogen}
                       &AC+AS+8 others &$1.49$ &    $\textbf{9.93}$    &  $1.82$   &  $2.63$   & $6.66$ & $4.53$  &  $40.30$   & $3.56$\\
AudioLDM~\cite{audioldm}
                       &AC+AS+2 others &$2.22$ &    $7.54$    &  $2.98$  & $2.48$     & $5.63$   & $5.65$ & $40.17$   & $3.08$ \\

\multirow{1}{*}{Tango2~\cite{tango}}
                       &AudioCaps & $1.32$   &  $9.12$   &  $2.03$   & $2.19$     & $6.84$   & $4.99$ & $43.39$   & $3.85$ \\
\multirow{1}{*}{AudioLDM2-Large~\cite{audioldm2}}
                       &AC+AS+6 others &  $\textbf{1.22}$   &  $7.86$   &  $1.83$  &  $1.65$     & $7.61$   &  $2.92$  &   $38.05$   & $3.47$ \\
\midrule
\multirow{1}{*}{AudioLDM-T5}
                       & Sound-VECaps$_{\operatorname{F}}$ &  $1.68$   &  $6.8$   &  $1.78$   &  $1.44$     &  $6.29$    &  $1.45$   & $41.20$    &$3.92$ \\
\multirow{1}{*}{AudioLDM-T5-L}
                       & Sound-VECaps$_{\operatorname{F}}$ &  $1.49$   &  $8.77$   &  $\textbf{1.49}$  &   $\textbf{1.17}$    &  $\textbf{7.96}$    &  $\textbf{1.06}$ &   $\textbf{43.59}$  & $\textbf{4.05}$ \\
\bottomrule
\end{tabular}
\label{tab:result}
\vspace{-0.5cm} % 减少0.5厘米的垂直空间
\end{table*}

\section{Audio Generation System}
\label{sec:method}

We conduct experiments on text-to-audio generation using AudioLDM~\cite{audioldm} models, a SoTA audio generation model, to evaluate Sound-VECaps. For instance, AudioLDM is divided into four sections: a CLAP encoder for condition embedding, a latent diffusion-based model to generate audio features within the latent space, a variational autoencoder~(VAE) decoder to reconstruct the information into a mel spectrogram, and a generative adversarial network(HiFi-GAN) vocoder~\cite{hifigan} to produce the waveform as the final output. 

Instead of using CLAP~\cite{clap} for computing the audio and text embedding, our experiments replaced the encoder with a T5~\cite{t5} encoder for condition embedding, and an cross-attention module~\cite{reaudioldm} is applied to process the T5 embedding. We name the system as AudioLDM-T5. For the remaining modules, we follow the same design of AudioLDM and our system takes the pre-trained VAE decoder and Hifi-GAN~\cite{hifigan} vocoder for audio feature reconstruction.

\section{Experiments}
\label{sec:exp}

% Various experiments are conducted to evaluate the effectiveness of using Sound-VECaps for audio generation and explore the usefulness of visual-only content. Additionally, the potential of the proposed dataset for other audio-language tasks is investigated.

\subsection{Evaluation Dataset}

We first follow previous baseline models~\cite{audioldm,audiogen} and evaluate the performance of text-to-audio generation on the AudioCaps testing set. Since AudioCaps only includes simple and audio-only textual information, we introduce a novel benchmark with enriched and enhanced captions~(same audio with better captions). We apply the proposed re-captioning pipeline in Section~\ref{sec:Dataset Construction} to generate improved captions for AudioCaps testing audio samples. Specifically, human supervision is applied during the captioning process to check the accuracy and relevance of each caption~(ensure the quality of LLM outputs). The proposed AudioCaps-Enhanced testing dataset includes five different captions for each audio clip, totalling $4430$ captions for $886$ audio samples. Similar to the Sound-VECaps dataset, we provide both full-feature captions~(AudioCaps-Enhanced$_{\operatorname{F}}$) and captions excluding visual-only contents~(AudioCaps-Enhanced$_{\operatorname{A}}$) for various purposes.

\subsection{Results}

\noindent\textbf{Effectiveness on Audio Generation.}
Audio generation systems are trained on Sound-VECaps to evaluate their effectiveness, where models are trained using the same hyperparameters of AudioLDM. Specifically, AudioLDM-T5 maintains the same size as AudioLDM~\cite{audioldm}, while AudioLDM-T5-L is a larger system with increased hidden sizes. As shown in Table~\ref{tab:result}, AudioLDM-T5 achieves SoTA performance on the AudioCaps testing sets. Moreover, the larger model~(AudioLDM-T5-L), trained on Sound-VECaps$_{\operatorname{F}}$, outperforms baseline models by a large margin. In addition, current audio generation models struggle with complex and extended prompts, resulting in performance degradation on AudioCaps-Enhanced (e.g., the FAD score increases from $1.83$ to $2.92$ on AudioLDM2-L). By applying Sound-VECaps for training, AudioLDM-T5 models successfully overcome this limitation, achieving a FAD score of $1.06$ and a MOS score of $4.05$ on larger AudioLDM-T5-L. 

\begin{table}[bp]
\vspace{-0.8cm} % 减少0.5厘米的垂直空间
\caption{AudioLDM-T5-L models trained and evaluated on different datasets, where E$_{\operatorname{F}}$ is short for Enhanced$_{\operatorname{F}}$ with full feature captions and AudioCap-E$_{\operatorname{A}}$ without visual-only contents. }
\centering
\small
\resizebox{0.475\textwidth}{!}{%
\begin{tabular}{ccccccc}
\toprule
 \multicolumn{1}{c}{\textbf{Training Dataset}} & \multicolumn{1}{c}{\textbf{Testing Dataset}}& \multicolumn{1}{c}{\textbf{KL}$\downarrow$}& \multicolumn{1}{c}{\textbf{IS}$\uparrow$}& \multicolumn{1}{c}{\textbf{FAD}$\downarrow$ } \\
\midrule
 Sound-VECaps$_{\operatorname{A}}$& AudioCaps&$1.22$ & $7.31$ & $1.65$ \\
 Sound-VECaps$_{\operatorname{A}}$& AudioCaps-E$_{\operatorname{F}}$&$1.33$& $6.27$ & $1.67$ \\
 Sound-VECaps$_{\operatorname{A}}$& AudioCaps-E$_{\operatorname{A}}$& $1.38$ & $7.18$ & $1.64$ \\
 \midrule
 Sound-VECaps$_{\operatorname{F}}$ & AudioCaps& $1.49$ & $\textbf{8.77}$ & $1.49$ \\
 Sound-VECaps$_{\operatorname{F}}$ & AudioCaps-E$_{\operatorname{F}}$& $\textbf{1.17}$ & $7.96$ & $1.06$ \\
 Sound-VECaps$_{\operatorname{F}}$ & AudioCaps-E$_{\operatorname{A}}$& $1.19$   & $8.13$ & $\textbf{0.96}$ \\
% \midrule

\bottomrule
\end{tabular}
}

\label{tab:filter experiment}
\end{table}

\begin{table*}[htbp]\small
\caption{Performance comparison between different systems on AudioCaps and AudioCaps-Enahcned(proposed benchmark), CLAP$_{M}$ and CLAP$_{L}$ are models trained by Microsoft~\cite{CLAP-micro} and LAION~\cite{clap}, using different structures and datasets respectively. For the training set, ``AC'', ``CL'' and ``LA'' are short for AudioCaps, Clotho and LAION-Audio-630k datasets respectively.}
\centering
\resizebox{1.0\textwidth}{!}{%
% \begin{tabular}{ccp{0.03325\linewidth}p{0.03325\linewidth}p{0.03325\linewidth}p{0.03325\linewidth}p{0.03325\linewidth}p{0.03325\linewidth}|p{0.03325\linewidth}p{0.03325\linewidth}p{0.03325\linewidth}p{0.03325\linewidth}p{0.03325\linewidth}p{0.03325\linewidth}p{0.03325\linewidth}p{0.03325\linewidth}}
\begin{tabular}{cccccccc:cccccc}
\toprule
\multirow{4}{*}{\textbf{Model}} & \multirow{4}{*}{\textbf{Training Set}} & \multicolumn{6}{c}{\textbf{AudioCaps}} & \multicolumn{6}{c}{\textbf{AudioCaps-Enhanced}}  \\
\cmidrule(lr){3-14}
& & \multicolumn{3}{c}{\textbf{Text-to-Audio}} & \multicolumn{3}{c}{\textbf{Audio-to-Text}} & \multicolumn{3}{c}{\textbf{Text-to-Audio}} & \multicolumn{3}{c}{\textbf{Audio-to-Text}} \\ 
\cmidrule(lr){3-14}
& & \textbf{R@1} & \textbf{R@5} & \textbf{R@10} & \textbf{R@1} & \textbf{R@5} & \multicolumn{1}{c}{\textbf{R@10}} & \multicolumn{1}{c}{\textbf{R@1}} & \textbf{R@5} & \textbf{R@10} & \textbf{R@1} & \textbf{R@5} & \textbf{R@10} \\ 
\midrule

CLAP$_{L}$~\cite{clap} & AC+CL+LA & $34.2$  & $71.1$ & $84.1$ & $43.1$ & $79.5$ & $90.1$ & $21.6$  & $54.9$ & $71.6$ & $34.1$ & $65.4$ & $77.7$ \\

CLAP$_{M}$~\cite{CLAP-micro} & 4.6M-Audio  & $33.5$  & $70.4$ & $80.2$ & $47.8$ & $80.2$ & $90.7$ & $19.5$  & $46.2$ & $60.9$ & $29.3$ & $59.1$ & $70.1$ \\

WavCaps~\cite{wavcaps} & WavCaps+AC+CL  & $39.7$  & $74.5$ & $86.1$ & $51.7$ & $82.3$ & $90.6$ & $23.0$  & $52.3$ & $66.2$ & $35.5$ & $62.8$ & $75.8$ \\

Auto-ACD~\cite{autoacd} & Auto-ACD  & $40.4$  & $\textbf{75.3}$ & $\textbf{87.4}$ & $51.1$ & $84.0$ & $92.7$ & $46.3$  & $81.8$ & $89.7$ & $55.8$ & $84.1$ & $92.6$ \\

\midrule
CLAP & Sound-VECaps-Audio & $\textbf{41.2}$  & $74.5$ & $85.3$ & $53.3$ & $83.2$ & $93.0$ & $49.2$  & $83.1$ & $91.7$ & $59.1$ & $87.5$ & $94.3$ \\
% \midrule

CLAP & Sound-VECaps-Full &  $39.2$&  $74.1$ & $85.0$ & $\textbf{54.0}$ & $\textbf{85.5}$ & $\textbf{93.2}$ & $\textbf{53.1}$&  $\textbf{85.7}$ & $\textbf{91.3}$ & $\textbf{64.3}$ & $\textbf{90.2}$ & $\textbf{96.4}$ \\
\bottomrule
\end{tabular}
}
\label{tab:retrieval}
\vspace{-0.4cm} % 减少0.5厘米的垂直空间
\end{table*}

\noindent \textbf{Effectiveness of Visual-Only Content.} 
To evaluate the effectiveness of the visual information in the captions, we compare the performance of different AudioLDM-T5-L systems trained and evaluated on various datasets that include and exclude visual-only content. Notably, all three versions of the testing dataset share the same group of audio clips~(same target audio samples while using different prompts for generation), providing reliability assurance for the comparison. As shown in Table~\ref{tab:filter experiment}, systems utilizing Sound-VECaps$_{\operatorname{F}}$ as the training dataset demonstrates enhanced performance across all three evaluation metrics. For the evaluation, using AudioCaps as the prompt presents a higher quality~(IS score of $8.77$), while the audio outputs generated through the prompts with visual content~(AudioCaps-Enhanced$_{\operatorname{F}}$) show minor degradation. However, audio samples generated through enriched prompts lead to significant improvements in the fidelity of generated audio, with the prompts excluding visual-only content~(AudioCaps-Enhanced$_{\operatorname{A}}$) showing SoTA performance. Through these experiments, we have summarized three key findings: 1) Training on captions with visual features can improve the capability of the system to handle auditory information and identify features across different modalities, leading to significant improvement in the overall performance. 2) The simplicity of the prompts in current evaluation benchmarks~(e.g. AudioCaps) limits the presentation of detailed audio features. The proposed benchmark testing on AudioCaps-Enhanced enriches the information with more controllable features and offers greater potential for enhancing the output quality. 3) Although training with external visual features~(Sound-VECaps$_{\operatorname{F}}$) provides better results, the additional visual information may increase data complexity during inference. Therefore, the system that uses prompts without visual-only features (AudioCaps-Enhanced$_{\operatorname{A}}$) generates the best result~($0.96$ on FAD).

\subsection{Studies on Other Audio Tasks}

\noindent \textbf{Audio Caption Retrieval.} 
\label{sec:retrieval}
In addition to our experiments on audio generation, we evaluated the effectiveness of Sound-VECaps for improving audio-language retrieval systems. Specifically, we employed the framework in WavCaps~\cite{wavcaps},~which uses RoBERTa as the text encoder and HTSAT as the audio encoder, to train and evaluate CLAP-based models in audio-text cross modal retrieval tasks. As illustrated by Table~\ref{tab:retrieval}, the evaluation using the AudioCaps testing set demonstrated that the CLAP-based models trained on the Sound-VECaps dataset matched the performance of the baseline models~(trained on other datasets). However, when testing with the enhanced captions~(AudioCaps-Enhanced), the experiment shows a notable performance decline in current SoTA systems, highlighting the challenges posed by longer and more detailed textual information. Conversely, the systems trained with enriched captions~(Auto-ACD~\cite{autoacd} and Sound-VECaps) present improvements in retrieval capabilities, while the system on Sound-VECaps$_{\operatorname{F}}$ achieves the best performance. The results show the enhancement of captions through visual information, as well as the accuracy and robustness of the system on Sound-VECaps. Additionally, the CLAP model trained with Sound-VECaps$_{\operatorname{F}}$ exhibited better performance, particularly on AudioCaps-Enhanced dataset, indicating that the overall performance of the system can be further improved with the visually augmented captions. 
% \noindent \textbf{Temporal-feature Retrieval:} Despite traditional retrieval tasks, we also examie the performance on identifiying the temporal feature by applying the evaluation pipeline of T-CLAP~\cite{tclap}

\noindent \textbf{Temporal Feature Retrieval.} 
Another aspect of Sound-VECaps is the temporal information. Since visual captions are provided by frame, temporal information~(events ordering) is also included. We applied the T-Classify method from T-CLAP~\cite{tclap} to evaluate the performance of temporal feature retrieval. Table~\ref{tab:order} demonstrates stronger capabilities to identify temporal information in systems with Sound-VECaps, illustrating its improvement in temporal features. In addition, the system developed without visual-only contents presents better performance, indicating that extensive visual features might influence the model's understanding of temporal information. 

\noindent \textbf{Limitation.} 
We also attempt to use Sound-VECaps for other audio-related tasks. However, due to the rich content in our captions, particularly regarding visual information, the model did not perform well on tasks that are purely audio-targeted content, such as audio captioning and zero-shot tasks. These results demonstrate that Sound-VECaps may not be broadly applied to audio-language tasks. It is mainly effective in a range of tasks that require processing and distinguishing detailed content, such as generation and retrieval.

\begin{table}[htbp]
\caption{Results of Temporal feature retrieval on T-Classify~\cite{tclap}.}
\centering
\small
\begin{tabular}{ccc}
\toprule

\multirow{1}{*}{\textbf{Model}} 
  & \multicolumn{1}{c}{\textbf{Text-to-Audio}} & \multicolumn{1}{c}{\textbf{Audio-to-Text}}  \\
\midrule
CLAP$_{m}$~\cite{CLAP-micro}  & $45.7$ & $44.1$ \\

CLAP$_{l}$~\cite{clap}  & $56.2$ & $53.2$ \\

WavCaps~\cite{wavcaps}  & $58.5$ & $49.7$ \\
\midrule
 Sound-VECaps$_{\operatorname{F}}$ & $61.2$ & $57.3$ \\

 Sound-VECaps$_{\operatorname{A}}$ & $\textbf{63.6}$ & $\textbf{59.0}$  \\
\bottomrule
\end{tabular}

\label{tab:order}
\vspace{-0.4cm} % 减少0.5厘米的垂直空间
\end{table}

\section{Conclusion}
\label{sec:conclusion} 

We present Sound-VECaps, a large-scale dataset comprising $1.66$M audio clips with captions augmented by video data, to address the challenge of prompt following in audio generation systems. Experiments show that systems trained on Sound-VECaps achieve SoTA performance and outperform baseline models. In addition, a new benchmark using improved captions is proposed to evaluate audio-language systems on complex and extended prompts. Our systems are further improved by a large margin when taking more detailed captions as prompts, reaching a FAD score of $0.96$. Nevertheless, we demonstrated that using Sound-VECaps can offer substantial improvements in audio-language and temporal feature retrieval. The results of the proposed AudioCaps-Enhanced testing sets highlight the limitations of previous benchmarks and emphasize the potential of better prompts in advancing the performance of audio-language models. Overall, we develop two versions of the proposed datasets with captions that include and exclude visual-only content for different purposes and tasks and hope these datasets will generate more profound impacts on audio-language learning.

\section{ACKNOWLEDGMENT}
\label{sec:ack}
This research was partly supported by a research scholarship from the China Scholarship Council~(CSC), an internship from ByteDance, and supported by the Engineering and Physical Sciences Research Council (EPSRC) under Grant EP/T019751/1 “AI for Sound”. For the purpose of open access, the authors have applied a Creative Commons Attribution~(CC BY) license to any Author Accepted Manuscript version arising.
% -------------------------------------------------------------------------
% \bibliographystyle{IEEEtran}
% \bibliography{strings,refs}

% 强制分页，并开始附录
\newpage
\appendix
\section{Appendix}
\label{sec:appendix}

\subsection{LLMs Prompts}
\begin{figure}[htbp]
    \centering
    \includegraphics[width=0.95\linewidth]{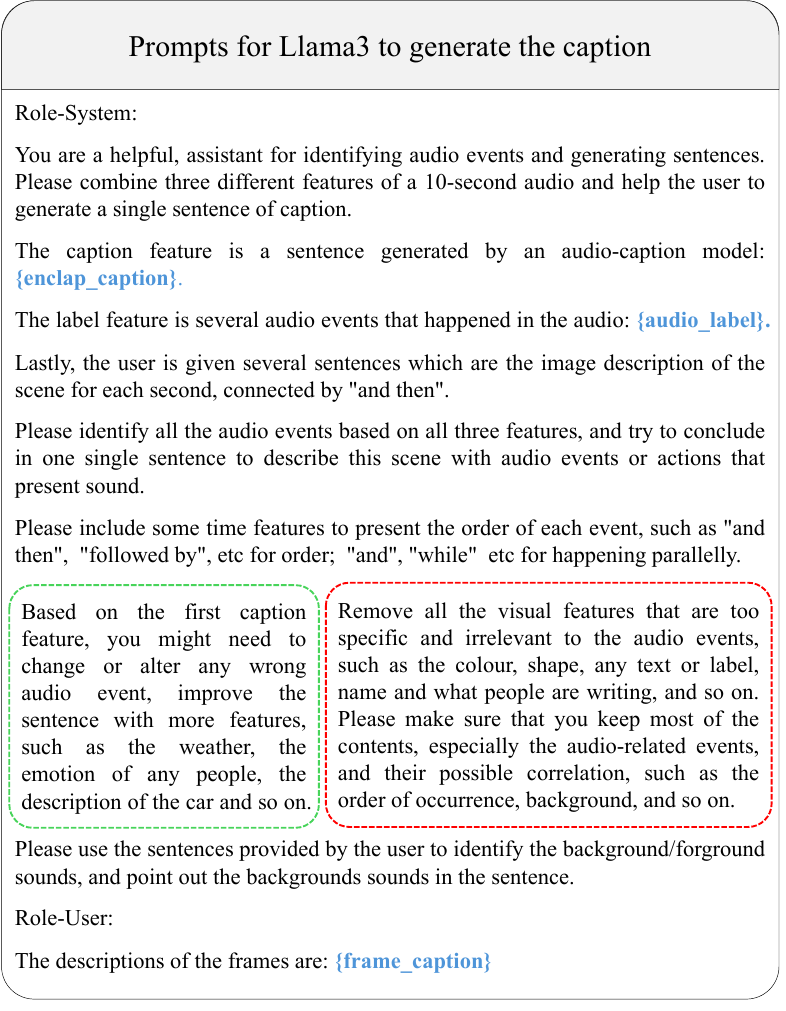}
    \caption{The prompts used for caption generation, where the contents in green section are used for full feature captions and red sections are applied to avoid any visual-only contents.}
    \label{fig:prompt}
\end{figure}

We provide the prompts used as the input for the Llama3 model to generate our proposed captions. As shown in the Figure~\ref{fig:prompt}, the prompt is a combination of three different features. In the system section, both the caption from Enclap and the audio label are provided, while the frame captions are presented as the user input. Two different contents are also provided for both the full-featured caption~(section in green boundaries) and the caption that filtered all the visual-only contents~(section in red boundaries).  For the AudioCaps-Enhanced dataset, we apply the same prompting pipeline, while changing the caption of enclap into the actual caption provided by the AudioCaps testing set. Nevertheless, all the captions for AudioCaps-Enhanced are generated under human-involved supervision, to ensure the correctness and relevance of the prompts.

\subsection{Caption Demos}
We present the comparison of the captions from Sound-VECaps and other baseline datasets. As shown in Table~\ref{tab:demo}, for each audio sample, we compare the caption from the AudioSet label, Wavcaps, Enclap, Auto-ACD and two versions of the proposed Sound-VECaps. A sample of the proposed AudioCaps-Enhanced testing dataset is presented in Table~\ref{tab:demo-2}.

\begin{table}[t]

\centering
\footnotesize
\caption{The comparison between different caption datasets.}
\renewcommand{\arraystretch}{1.05}
\begin{tabular}{p{0.5cm}cp{5.7cm}}
\toprule

\multirow{1}{*}{\textbf{Num}} 
  & \multicolumn{1}{c}{\textbf{Dataset}} & \multicolumn{1}{c}{\textbf{Caption}}  \\

\midrule
\multirow{13}{*}{No.1} &AudioSet & Honk , Speech \\

% \midrule
& WavCaps  & Crinkling, wind, laughter, ducks, and people speaking are heard. \\

& Enclap & Wind blows, ducks quack and people speak. \\

& Auto-ACD & The wind blows as ducks quack and a man speaks. \\

% \midrule
& \multirow{4}{*}{VECaps$_{\operatorname{a}}$} &  A goose quacks and honks, while the wind blows, and the person speaks, followed by the sound of bread being offered to the goose, amidst the scattered leaves and grass. \\

& \multirow{6}{*}{VECaps$_{\operatorname{f}}$} &  As the person stands near the car, a goose quacks and honks, while the wind blows, and the person speaks, followed by the sound of bread being offered to the goose, and the goose's orange beak and feet can be seen amidst the scattered leaves and grass.  \\
\midrule
\multirow{12}{*}{No.2} &AudioSet & Dial tone \\

% \midrule

& WavCaps  & A dial tone is heard. \\

& Enclap & A telephone rings \\

& Auto-ACD & A dial tone rings with a probability of 0.66, indicating a telephone call in an indoor setting. \\

% \midrule
& \multirow{4}{*}{VECaps$_{\operatorname{a}}$} & A telephone rings in the background, followed by a dial tone, while a man is holding a child in his arms, as a news article plays in the background. \\

& \multirow{6}{*}{VECaps$_{\operatorname{f}}$} & A telephone rings in the background, followed by a dial tone, while a man is holding a child in his arms in front of a destroyed building, as a news article about US urging Israel to protect civilians and increase aid to Gaza plays in the background.  \\
\midrule
\multirow{15}{*}{No.3} &AudioSet & Music, instrument,  string \\

% \midrule

& WavCaps  & Music is playing. \\

& Enclap & A man speaks over a loudspeaker as music plays in the distance \\

& Auto-ACD & The sitar player strums melodious music on stage, accompanied by instruments in an orchestra pit. \\
% \midrule
& \multirow{4}{*}{VECaps$_{\operatorname{a}}$} & A man plays a sitar, accompanied by the sound of a plucked string instrument, followed by the soft hum of a bowed string instrument, in a dimly lit room, with music playing in the distance. \\

& \multirow{6}{*}{VECaps$_{\operatorname{f}}$} & A man plays the sitar, a traditional Indian stringed instrument, in a dimly lit room with a projection screen in the background, while music plays in the distance, accompanied by the sound of a plucked string instrument, followed by the soft hum of a bowed string instrument.  \\

\bottomrule
\end{tabular}

\label{tab:demo}
\end{table}

\begin{table}[tbp]

\centering
\footnotesize
\caption{The comparison between AudioCaps and proposed AudioCaps-E testing dataset.}
\renewcommand{\arraystretch}{1.1}
\begin{tabular}{cp{5.7cm}}
\toprule

\multicolumn{1}{c}{\textbf{Dataset}} & \multicolumn{1}{c}{\textbf{Caption}}  \\

\midrule
AudioCaps  & A man talking as water splashes. \\

% \midrule
\multirow{4}{*}{AudioCaps-E$_{\operatorname{a}}$} & Waves crashing onto a calm shore, followed by a man speaking amid a gathering of people, some with cameras, by a coastal backdrop. \\

\multirow{4}{*}{AudioCaps-E$_{\operatorname{f}}$} &  Waves gently lap against the shore under an overcast sky, as a man in a grey shirt and glasses addresses a gathering. Surrounding him, a few individuals, possibly security or journalists, hold cameras and microphones, suggesting a public event near a tropical waterfront.  \\

\bottomrule
\end{tabular}

\label{tab:demo-2}
\end{table}

\end{document}